\documentclass[iop]{emulateapj}

\usepackage{graphicx}
\usepackage{verbatim}
\slugcomment{\it{}Accepted for publication in the Astrophysical Journal Letters, July 21, 2015}

\newcommand{\Fermi}{{\it{}Fermi}\ }
\newcommand{\n}{\nodata}
\newcommand{\be}{\begin{itemize}}
\newcommand{\ee}{\end{itemize}}

\def\fermi{\textit{Fermi }}

\makeindex
\citeindextrue

\shorttitle{Why Haven't Many of the Brightest Radio Loud Blazars Been
  Detected by {\it Fermi} ?}
\shortauthors{M. L. Lister et al.}

\begin{document}

\title{Why Haven't Many of the Brightest Radio Loud Blazars Been
  Detected in Gamma-Rays by {\it Fermi} ?}

\author{ M. L. Lister\altaffilmark{1},
M. F. Aller\altaffilmark{2},
H. D. Aller\altaffilmark{2},
T. Hovatta\altaffilmark{3,4},
W. Max-Morbeck\altaffilmark{5},
A. C. S. Readhead\altaffilmark{4},
J. L. Richards\altaffilmark{1},
E. Ros\altaffilmark{6,7,8}
}

\altaffiltext{1}{
Department of Physics, Purdue University, 525 Northwestern Avenue,
West Lafayette, IN 47907, USA;
\email{mlister@purdue.edu}
}

\altaffiltext{2}{ 
  Department of Astronomy, University of Michigan, 311 West Hall, 1085
  S. University Avenue, Ann Arbor, MI 48109, USA; }

\altaffiltext{3}{Aalto University, Mets\"ahovi Radio Observatory, 
  Mets\"ahovintie 114, FI-02540, Kylm\"al\"a, Finland;}

\altaffiltext{4}{
Cahill Center for Astronomy \& Astrophysics, California Institute of
Technology, 1200 E. California Blvd, Pasadena, CA 91125, USA;}

\altaffiltext{5}{
National Radio Astronomy Observatory, P.O. Box O 1003 Lopezville Road
Socorro, NM 87801-038, USA;} 

\altaffiltext{6}{
Max-Planck-Institut f\"ur Radioastronomie, Auf dem H\"ugel 69,
53121 Bonn, Germany}

\altaffiltext{7}{
Observatori Astron\`omic, Universitat de Val\`encia,
  Parc Cient\'{\i}fic, C. Catedr\'atico Jos\'e Beltr\'an 2, E-46980
  Paterna, Val\`encia, Spain;}

\altaffiltext{8}{
Departament d'Astronomia i Astrof\'{\i}sica,
  Universitat de Val\`encia, C. Dr. Moliner 50, E-46100 Burjassot,
  Val\`encia, Spain;}

\begin{abstract}
  We use the complete MOJAVE 1.5 Jy sample of active galactic nuclei
  (AGN) to examine the gamma-ray detection statistics of the brightest
  radio-loud blazars in the northern sky.  We find that 23\% of these
  AGN were not detected above 0.1 GeV by the {\it Fermi} LAT during
  the 4-year 3FGL catalog period partly because of an instrumental
  selection effect, and partly due to their lower Doppler boosting
  factors. Blazars with synchrotron peaks in their spectral energy
  distributions located below $10^{13.4}$ Hz also tend to have
  high-energy peaks that lie below the 0.1 GeV threshold of the LAT,
  and are thus less likely to be detected by {\it Fermi}. The
  non-detected AGN in the 1.5 Jy sample also have significantly lower 15 GHz
  radio modulation indices and apparent jet speeds, indicating that
  they have lower than average Doppler factors. Since the
  effective amount of relativistic Doppler boosting is enhanced in
  gamma-rays (particularly in the case of external inverse-Compton
  scattering), this makes them less likely to appear in the 3FGL
  catalog. Based on their observed properties, we have identified
  several bright radio-selected blazars that are strong candidates for
  future detection by {\it Fermi}.

\end{abstract}
 
$\;$

\section{INTRODUCTION} 

The {\it Fermi} space telescope is a powerful broad-band gamma-ray
facility that has continuously scanned the entire sky every 3 hours
since 2008. One of its major discoveries has been that away from the
galactic plane, the gamma-ray sky is dominated by the blazar class of
active galactic nuclei (AGN) \citep{1FGL}. These relatively rare AGN
harbor powerful jets of relativistically-moving plasma that are
oriented close to our line of sight. Their overall spectral energy
distribution (SED) tends to be dominated by relativistically boosted
emission from the jet, and typically consists of two broad peaks in a
plot of log $\nu F_\nu$ versus log $\nu$, where $F_\nu$ is the
observed flux density at frequency $\nu$ \citep{abdo_sed}. The lower
frequency peak is associated with synchrotron emission from
relativistic electrons in the jet plasma, while the high frequency
peak is widely believed to be created by inverse-Compton (IC)
up-scattering of photons from the jet, accretion disk and/or broad line
region \citep[e.g.,][]{1993ApJ...416..458D,1994ApJ...421..153S}.

Since the same population of jet electrons is responsible for these
two peaks, we might expect the synchrotron and gamma-ray emission from
individual blazars to be correlated, and this has proven to be the
case. Many studies have found  statistical correlations between the
radio/sub-mm band and the $>0.1$ GeV time-averaged fluxes measured by
the {\it Fermi} LAT instrument \citep[e.g.,][]{2009ApJ...696L..17K,
  2011ApJ...741...30A,2011ApJ...742...27L,2012AA...541A.160G,2014MNRAS.441.1899F,2015MNRAS.450.2658M}.
\cite{2010ApJ...718..587M} have also shown that the fraction of
radio-loud AGN detected by \fermi steadily increases with increasing 20
GHz flux density.

Despite these well-established correlations, it is still not fully
understood why a substantial fraction of the brightest, most compact
blazars in the GHz band have not been detected at energies above 0.1
GeV, despite the four years of continuous observations included in the
most recent \fermi catalog \citep[3FGL;][]{3FGL}. In this {\it
  Letter}, we investigate the properties of a complete flux
density-limited, 15 GHz AGN sample of the northern sky, the MOJAVE 1.5
Jy survey, where 23\% of the AGN have no {\it Fermi} detections. We
show that these non-detected AGN have a lower than average jet speed
and radio variability index (indicative of a lower Doppler boosting
factor) and a synchrotron component that peaks below $\sim 10^{13.4}$
Hz. The latter causes their secondary SED peak to be located well
below the lower energy cutoff of the LAT instrument, resulting in a
low LAT gamma-ray flux and non-detection by {\it Fermi}.  We discuss
the 1.5 Jy sample and observational data in \S~2, present our analysis
in \S~3, and summarize our conclusions in \S~4.
 
\section{\label{obsdata}OBSERVATIONAL DATA}

In order to properly evaluate the statistics of gamma-ray detections
of bright radio loud blazars, it is essential to use well-defined,
complete samples.  The GeV and radio bands above $\sim 10$ GHz are
ideal in this respect due to the lack of foreground obscuration (away
from the galactic plane), and little contamination from non-active
galaxies. In these bands, the AGN emission is dominated by that of the
jet, with little contribution from the extended lobes.  In this
section we describe the radio-selected MOJAVE 1.5 Jy AGN sample, and
the supporting data from our own studies and the literature.

\subsection{Radio data} 

In the GHz band, there have been many large sky surveys carried out
over the last few decades, e.g., FIRST \citep{1995ApJ...450..559B},
NVSS \citep{1998AJ....115.1693C}, GB6 \citep{1996ApJS..103..427G},
AT20G \citep{2010MNRAS.402.2403M}, such that virtually every radio
loud AGN above 1 Jy has been identified and cataloged. Nevertheless, a
major challenge in obtaining complete samples of radio loud blazars
has been their highly variable flux densities, which over the course
of several years can change by factors exceeding $\sim 10$ in the GHz
band \citep{2014MNRAS.438.3058R}, and $\sim 200$ in the GeV band
\citep{2010ApJ...722..520A, 2014MNRAS.441.1899F}.  Any single epoch
flux-limited survey therefore stands to miss a substantial
portion of the population.

As part of a VLBA key project to study the parsec-scale structure and
evolution of AGN jets \citep{MOJAVE_V}, we have constructed a
complete, flux density-limited sample, based on all available GHz band flux
density data on bright radio AGNs over a 16 year period from 1994.0 to
2010.0 \citep{MOJAVE_X}. The MOJAVE 1.5 Jy sample includes all AGN
(excluding gravitational lenses) located north of J2000 declination
$-30^\circ$ that are known to have exceeded 1.5 Jy in compact
(milliarcsecond-scale) 15 GHz flux density at least once during that period.
We were able to estimate the milliarcsecond-scale flux density from single-dish
measurements by using near-simultaneous VLBA/single-dish observations
of each source at several epochs. The difference of these
near-simultaneous measurements represents the amount of
arcsecond-scale emission resolved out by the VLBA, which is expected
to be non-variable due to its size.  Given the measurement errors, we
were able to detect any extended emission above 0.02 Jy.

Using newly obtained data from the OVRO 15 GHz AGN monitoring program
\citep{2011ApJS..194...29R}, we have refined our extended flux density
estimates and subsequently dropped two AGN (MG1 J021114$+$1051 and OP
$-$050) from the original 1.5 Jy sample list.  We list the properties
of all 181 AGN in the revised sample in Table~1.  For the purposes of
this paper, we consider only the 163 AGN that are located at least 10
degrees away from the galactic plane. This ensures that issues related
to galactic gamma-ray foreground subtraction do not affect the LAT
detection statistics of our sample.

\subsection{Gamma-ray data}
The Third \fermi Gamma-Ray Catalog \citep[3FGL;][]{3FGL} is based on
LAT data collected between 2008 August 4 and 2012 July 31, and
contains 3033 high-confidence detections above 0.1 GeV. The Third LAT
AGN Catalog \citep[3LAC;][]{2015arXiv150106054A}, associates 1563 of
these sources with AGN.  Because of the relatively large sky position
errors ($\sim$ several arcmin) of the 3FGL sources, true counterpart
identification has only been achieved in $12\%$ of the cases. The
remainder of the 3FGL associations were established using extensive
statistical likelihood tests and a variety of celestial source
catalogs. The 3LAC list consists of only high-confidence 3FGL AGN
associations. For brevity, we will refer to these AGN as
``LAT-detected'' throughout this paper. Although a large number (992)
of 3FGL gamma-ray sources remain unassociated, virtually all of the
brightest ones have associations \citep[see,
e.g.,][]{3FGL,2011ApJ...742...27L}.

A total of 122 AGN in our sample are listed as associations in the
3LAC.  Additionally, PKS 0539$-$057 and 4C $+$ 06.69 appeared in the
1FGL catalog \citep{1FGL}, and the LAT detection of 3C~120 was announced by
\cite{2010ApJ...720..912A}.  The latter three AGN likely do not appear in
the 3FGL catalog since they experienced at most only a brief period of
gamma-ray flaring; thus their 4 year averaged fluxes were below the
3FGL significance cutoff.  We checked the remaining non-LAT detected
AGN in our sample, and in all cases the closest 3FGL source was at
least 17 arcminutes away.  There are also no \fermi associations for
any of these AGN in the refined list published by
\cite{2015ApJS..217....2M}.

\subsection{SED peak data} 

As discussed previously by \cite{2013ApJ...763..134F} there are
considerable inconsistencies in the SED synchrotron peak locations of
blazars reported in the literature. Since the peaks for most of our
sample lie between $10^{12}$ and $10^{14}$ Hz, where there is
typically a paucity of observational data, considerable
interpolation-based errors are present. Since our AGN are highly
core-dominated, any contribution from extended lobe emission to the SED is
negligible above $\sim 10^{10}$ Hz, however, some authors have included the
(non-jet) IR and optical big blue bump components in their polynomial
fits to the synchrotron SED component, resulting in overestimates of
the true peak location.

In order to minimize these errors, we have used the ASDC SED builder
Version 3.1.6 tool\footnote{https://tools.asdc.asi.it/} to estimate
the peak location using a 3rd degree polynomial fit to flux densities
published in the literature. For most of the AGN in our sample there
were hundreds of available measurements at numerous frequencies,
however, we only considered data up to $10^{14}$ Hz, and extended this
up to $10^{15}$ Hz only for clear cases where the SED remained
parabolic and there was no significant blue bump component. In some
instances, there were insufficient flux density data or too much
confusion from non-jet emission in the $10^{11.5}$ to $10^{14}$ Hz
region to reliably estimate the location of the synchrotron peak.
Based on comparisons to other published values in the literature for
our AGN, we estimate a typical error of $\pm 0.3$ dex in our values.

\section{\label{discussion}DISCUSSION}
\subsection{\label{syncpeak}\Fermi detection and synchrotron peak location}

The passband of the LAT instrument covers photon energies in the
approximate range of 100 MeV to 300 GeV \citep{2009ApJ...697.1071A}.
At the highest energies, AGN detections are typically limited by low
photon fluxes, since they typically have steep gamma-ray spectra with
power law energy indices ranging from $-$1.5 to $-$3 \citep{3FGL}. The
LAT's fixed passband tends to discriminate against AGN with IC peaks
located below 1 GeV, and those with very steep gamma-ray spectra. 
If the IC peak shapes of blazars are similar, then the LAT's fixed
passband should create a correlation between the LAT energy index and
the IC peak location. Unfortunately the latter are often difficult to
measure, due to a lack of observational data in the soft
gamma-ray and hard X-ray bands.  However, we would also expect the
locations of the synchrotron and IC peaks to be correlated, since
the same population of jet electrons is responsible for both features.
This has been shown to be the case, in the form of a relatively tight
correlation between LAT energy index and synchrotron peak location
\citep{1LAC}.  Therefore, since the two SED peaks track each other,
AGN with low synchrotron peak locations should have IC peaks located
well below the LAT bandpass, and consequently a smaller chance of
detection by {\it Fermi}.

\begin{figure}
\includegraphics[clip,angle=0,scale=0.36]{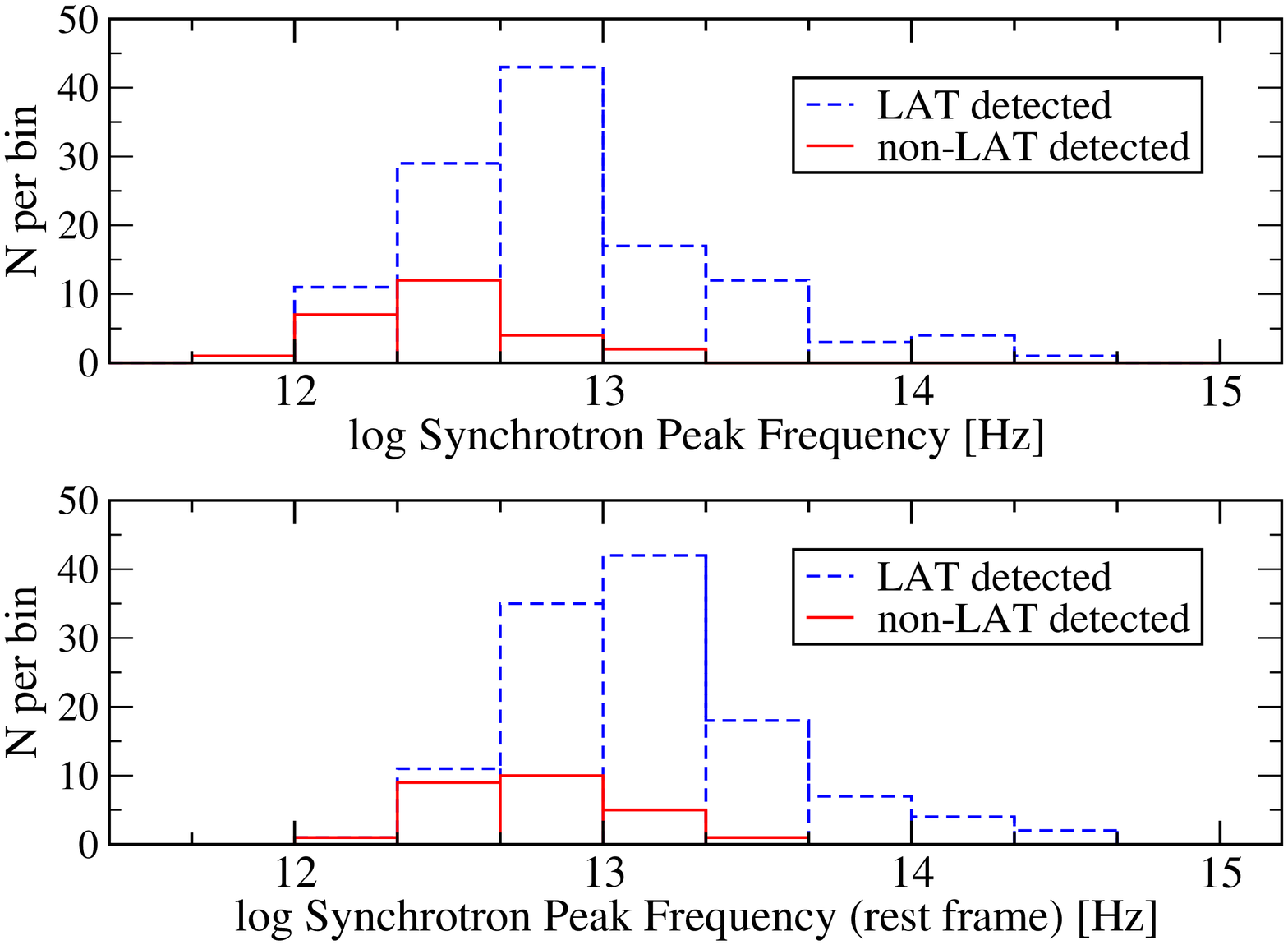}
\caption{\label{detectionhist}\footnotesize Top panel: distributions
  of observed synchrotron
peak frequency for LAT detected (blue dashed line) and non-LAT
detected (red solid line) sub-samples. Lower panel: same, but
for rest frame synchrotron peak frequency.}
\end{figure}

\begin{figure*}
\begin{center}
\plottwo{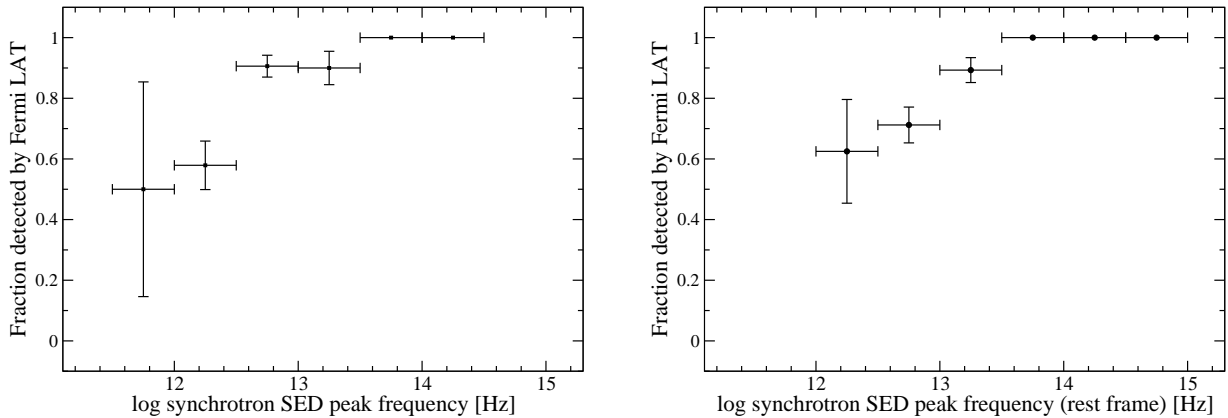}{LATdetection_vs_sedpeak_restframe.eps}

\end{center}
\caption{\label{detectionfraction}\footnotesize Left panel: LAT detection fraction versus
synchrotron peak frequency. The vertical error bar for each bin assumes a
binomial distribution with variance equal to $N\times f \times(f-1)$,
where $N$ is the number of AGN in the bin and $f$ is the LAT detection
fraction. Right panel: same, but for rest frame peak frequency.}
\end{figure*}

In Figure~\ref{detectionhist} we show histograms of synchrotron peak
frequency for the LAT and non-LAT detected sub-samples. All of
the non-detected AGN have synchrotron peak frequencies below
$10^{13.4}$ Hz (in both rest and observer frames). We performed two
commonly-used non-parametric statistical tests to look for differences
in the distributions.  The Kolmogorov-Smirnov test evaluates possible
differences in the cumulative distribution functions, while the
Wilcoxon rank-sum test evaluates whether the data points of one sample
are generally higher-valued (or lower-valued) than the other. We list
the test results in Table~2.  The LAT and non-LAT AGN have
significantly different synchrotron peak distributions, both in the
observed and rest frames.  Figure~\ref{detectionfraction} shows that
as predicted, the AGN in our sample with higher synchrotron peak
frequencies are more likely to be detected by {\it Fermi}.

\subsection{\label{jetspeed}\Fermi detection and Doppler boosting Indicators}

A main prediction of the external seed photon model for high-energy
blazar emission is that the IC radiation should be more highly Doppler
boosted than the synchrotron emission, due to a blue-shifting of the
external photons in the rest frame of the jet electrons
\citep{Dermer1995}. The gamma-rays will also experience more effective
boosting (via a $k$-correction) since their emission spectrum is much
steeper than that of the flat-spectrum radio jet. There is already
ample evidence that {\it Fermi} preferentially detects highly Doppler
boosted jets; the very small number of misaligned (low Doppler factor)
jets in the 3LAC (only 2\% of the catalog) is a prime example.  Other
studies showing this \fermi selection bias include those by
\cite{2009ApJ...696L..17K}, \cite{1LAC}, and \cite{2010AA...512A..24S}.

The Doppler factors of AGN jets are difficult to measure accurately,
due to the featureless power-law nature of the jet emission, and a
lack of a good means with which to estimate the jet viewing
angle. Some studies
\citep[e.g.,][]{1999ApJ...521..493L,2005AJ....130.1418J,Hovatta09}
have used light-crossing time arguments and variability timescales of
individual flares to derive Doppler factors, but only for a relatively
small number of AGN.  However, the overall variability level of a
blazar is also a good indicator of its Doppler factor, since any
intrinsic variability will be significantly shortened and increased in
amplitude by Doppler boosting
\citep[e.g.,][]{2001ApJ...561..676L}. The OVRO monitoring program has
published 15 GHz modulation indices for nearly 1500 blazars, covering
the time period 2008 January 1 to 2011 December 31. These indices are
estimates of the standard deviation of the flux density of a source
divided by its mean \citep{2011ApJS..194...29R}.

The gamma-ray detected AGN in the OVRO sample are significantly more
radio variable than the non-detected ones \citep{2014MNRAS.438.3058R},
and we find this same trend for the 142 OVRO AGN in our sample.
\cite{2014MNRAS.438.3058R} also found the level of AGN radio
variability to {\it decrease} with increasing synchrotron peak
frequency, but their sample includes significant numbers of AGN with
observed synchrotron peaks above $10^{14}$ Hz, whereas there are only
4 such AGN in our sample.

Another method of estimating the amount of relativistic boosting comes
from measurements of apparent jet speeds.  Although it is possible for
a jet to have a low apparent speed if it is aligned exceedingly close
to the line of sight (i.e., a viewing angle $< 1/\Gamma$, where
$\Gamma$ is the bulk Lorentz factor), for large flux-limited jet
samples the apparent speed will on average be well-correlated with
Lorentz factor \citep{2001ApJ...554..964L}. The MOJAVE program has
published maximum apparent jet speeds for 133 AGNs in our sample
\citep{MOJAVE_X}, and is currently obtaining multi-epoch VLBA data on
the remainder. 

In an earlier study using the first three months of {\it Fermi} data,
we showed that the gamma-ray detected AGN in the original MOJAVE
sample had higher apparent jet speeds \citep{2009ApJ...696L..22L}. This is also the
case for our 1.5 Jy sample, at an even higher level of statistical
confidence (Table~\ref{stattests}). Taken together with the trend in
modulation index, this indicates that the Doppler factor also plays an
important role in determining which radio-loud blazars will be
detected by {\it Fermi}.

\subsection{\Fermi detection and other jet properties}

In addition to Doppler factor and synchrotron peak location, there are
potentially other properties that can also influence the gamma-ray
flux of a blazar.  These include redshift, intrinsic jet luminosity,
and the activity state of the jet during the 3FGL observation period.
The latter is particularly important when considering the \fermi
detection statistics of the 1.5 Jy sample, which was selected over a
time period that only partially overlaps that of the 3FGL catalog.

We have compiled maximum and median 15 GHz VLBA flux
densities of the 1.5 Jy AGN during two time periods: a) between the
start of the 1.5 Jy sample selection window (1994 January 1) and the
start of \fermi operations (2008 August 4), and b) during the 3FGL
observation window.  We use the ratio of the maximum
flux density in these two periods as a {\it Fermi}-era radio activity
indicator, as was done by \cite{2009ApJ...696L..17K} in their analysis
of the \fermi LBAS list \citep{2009ApJS..183...46A}.
There were 51 AGN with insufficiently sampled flux density data in the
pre-\fermi era to determine a reliable activity index.

The OVRO program does not include any AGN below declination
$-20^\circ$, and ten other 1.5 Jy sample AGN were only  added to its
monitoring list after the end of the 3FGL period. In all of these
cases we had at least 4 epochs of
VLBA\footnote{http://www.astro.purdue.edu/MOJAVE} or UMRAO 14.5 GHz
\citep{2014ApJ...791...53A} observations during the 3FGL period with
which to calculate maximum and median values.

We find no statistically significant differences in the activity
indices of the LAT and non-LAT sub-samples. This is also the case for
the median radio flux density during the 3FGL time period. There is a
mild indication that the maximum radio flux densities of
the LAT-detected AGN are higher on average than the
non-detected ones. This is likely because the
majority of the LAT detected AGN are highly variable in gamma-rays,
and were detectable by {\it Fermi} only during a small
fraction of the 3FGL time window \citep{3FGL}.

There is no significant dependence of AGN LAT detection on redshift,
maximum 3FGL-era radio luminosity, or median 3FGL-era radio
luminosity. Our sample, being selected on radio jet emission,
is dominated by high-luminosity, high-redshift jets, and contains only
11 AGN located closer than redshift 0.1. Consequently, it does not uniformly
sample a wide range of jet luminosity, and is not ideally suited for
probing the gamma-ray detection statistics of lower luminosity AGN.

\section{\label{conclusions}Summary and Conclusions}

We have examined the \fermi LAT detection statistics during the time
period covered by the 3FGL catalog (2008 August 4 to 2012 July 31) of
the MOJAVE 1.5 Jy complete flux density-limited sample of radio-loud
blazars with J2000 declination $> -30^\circ$ and $|b| > 10^\circ$.  We
conclude that 23\% of these AGN were not detected above 0.1 GeV by the
\fermi LAT in part due to an instrumental selection effect, and partly
due to lower relativistic boosting of their jet emission.

\begin{figure*}
\begin{centering}
\plottwo{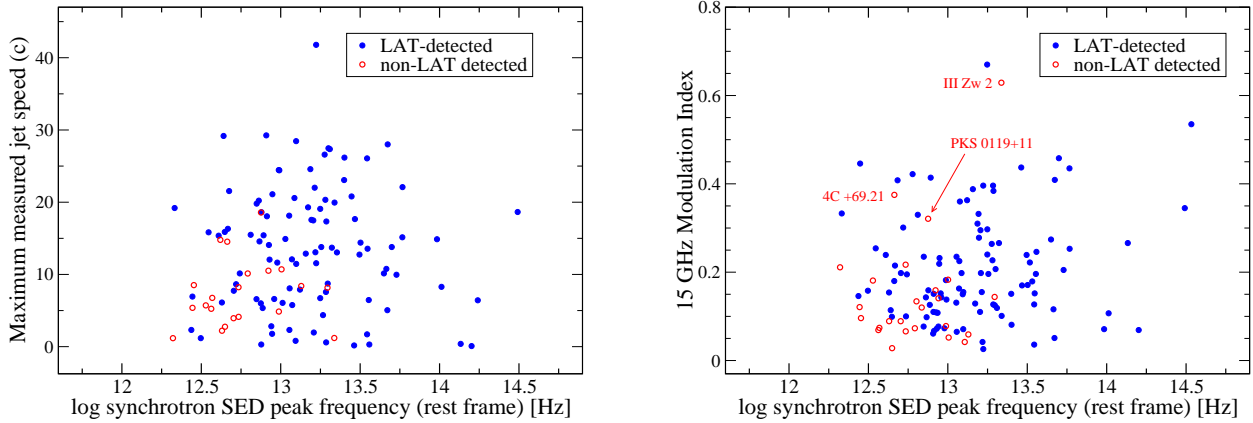}{modulation_vs_sedpeak_restframe.eps}
\end{centering}
\caption{\label{speedvssedpeak}\footnotesize Left panel: plot of maximum measured jet speed versus rest
frame synchrotron SED peak frequency. The blue filled circles represent
LAT-detected AGN, while the open red circles represent non-LAT
detected AGN. Right panel: same, but for 15 GHz OVRO modulation
index. }
\end{figure*}

The LAT-detected blazars have significantly higher radio variability
levels and apparent jet speeds than the non-detected ones.  Both of
these properties are positively correlated with the Doppler boosting
factor of the radio emission.  Since the effective amount of
relativistic flux boosting is enhanced in gamma-rays (especially in
the case of the external IC model), radio-selected blazars with lower
than average Doppler factors are less likely to be detected by {\it
  Fermi}.  In Figure~\ref{speedvssedpeak}, we plot apparent jet speed
and modulation index, respectively, versus the rest frame synchrotron
peak frequency.  In both scatterplots there is a clear tendency for
the non-LAT detected AGN to cluster in the lower left corner.  In
Figure~\ref{speedvssedpeak} we have identified three AGN (III Zw 2,
PKS 0119$+$11, and 4C $+$69.21), which based on their locations in the
plots, are likely candidates for future LAT detection. The LAT's
selection effects are clearly evident in the differences in the SED
synchrotron peak locations of the detected and non-detected blazars in
our sample.  Bright radio loud blazars with synchrotron peaks in their
SEDs located below $10^{13.4}$ Hz have high-energy peaks that lie well
below the lower energy cutoff of the LAT, resulting in a low gamma-ray
flux.  Future space telescopes such as {\it Astro-H} which cover the
hard X-ray/soft gamma-ray regime below 100 MeV therefore offer great
potential for detecting these blazars.

\acknowledgments 

The authors thank Ken Kellermann for helpful discussions.  The MOJAVE
program is supported under NASA-Fermi grant NNX12A087G.  MFA 
was supported by NASA-{\it Fermi} GI grants NNX09AU16G, NNX10AP16G
and NNX11AO13G and NSF grant AST-0607523. TH was supported by the
Academy of Finland project number 267324.  The National Radio
Astronomy Observatory is a facility of the National Science Foundation
operated under cooperative agreement by Associated Universities, Inc.
This work made use of the Swinburne University of Technology software
correlator \citep{2011PASP..123..275D}, developed as part of the
Australian Major National Research Facilities Programme and operated
under licence.

\ 
\clearpage
\vfill\eject
\LongTables
\begin{deluxetable*}{llllrrrccc} 
\label{gentable} 
\tablecolumns{10} 
\tabletypesize{\scriptsize} 
\tablewidth{0pt}  
\tablecaption{1.5 Jy MOJAVE AGN Sample Properties}  
\tablehead{\colhead{J2000} & \colhead {Alias} &\colhead{{\it Fermi} Catalog Name} &  
\colhead{z} &  \colhead{$S_\mathrm{ext}$} & \colhead{$S_\mathrm{max}$}& \colhead{$S_\mathrm{med}$} &\colhead{$a$} &   \colhead{$\nu_\mathrm{p,obs}$}& \colhead{$\nu_\mathrm{p,rest}$}  \\ 
\colhead{(1)} & \colhead{(2)} & \colhead{(3)} & \colhead{(4)} & \colhead{(5)} & 
 \colhead{(6)} & \colhead{(7)}  & \colhead{(8)}  & \colhead{(9)}  & \colhead{(10)}  } 
\startdata 
J0006$-$0623 & NRAO 005&\n & 0.3467& 0.08 & 2.69 & 2.09 & 0.78 & 13.0 & 13.1  \\ 
J0010+1058 & III Zw 2&\n & 0.0893& \n & 1.82 & 0.72 & 0.81 & 13.3 & 13.3  \\ 
J0019+7327 & S5 0016+73&\n & 1.781& \n & 1.48 & 1.10 & 0.66 & 12.3 & 12.7  \\ 
J0050$-$0929 & PKS 0048$-$09&3FGL J0050.6$-$0929 & 0.635& 0.04 & 2.33 & 0.82 & 1.10 & 14.3 & 14.5  \\ 
J0102+5824\tablenotemark{a} & TXS 0059+581&3FGL J0102.8+5825 & 0.644& \n & 5.41 & 2.49 & 1.21 & 12.7 & 12.9  \\ 
J0108+0135 & 4C +01.02&3FGL J0108.7+0134 & 2.099& 0.14 & 4.27 & 3.38 & 1.28 & 12.5 & 13.0  \\ 
J0112+2244 & S2 0109+22&3FGL J0112.1+2245 & 0.265& \n & 1.36 & 0.37 & 0.90 & 13.4 & 13.5  \\ 
J0112+3522 & B2 0109+35&\n & 0.450& \n & 1.37 & 0.99 & 0.89 & 12.8 & 13.0  \\ 
J0121+1149 & PKS 0119+11&\n & 0.570& 0.05 & 4.36 & 2.10 & 1.07 & 12.7 & 12.9  \\ 
J0125$-$0005 & UM 321&\n & 1.0765& 0.07 & 1.07 & 0.89 & 0.66 & \n & \n  \\ 
J0132$-$1654 & OC $-$150&3FGL J0132.6$-$1655 & 1.020& \n & 2.91 & 2.11 & 2.82 & 12.9 & 13.2  \\ 
J0136+4751 & DA 55&3FGL J0137.0+4752 & 0.859& \n & 4.90 & 3.31 & 0.86 & 12.3 & 12.6  \\ 
J0204+1514 & 4C +15.05&3FGL J0205.0+1510 & 0.405& 0.24 & 1.26 & 0.82 & 0.42 & 12.5 & 12.6  \\ 
J0205+3212 & B2 0202+31&3FGL J0204.8+3212 & 1.466& \n & 3.51 & 2.53 & 1.05 & 12.3 & 12.7  \\ 
J0204$-$1701 & PKS 0202$-$17&3FGL J0205.2$-$1700 & 1.739& 0.12 & 1.46 & 1.31 & 1.03 & 12.5 & 12.9  \\ 
J0217+7349 & S5 0212+73&3FGL J0217.5+7349 & 2.367& 0.04 & 3.96 & 3.23 & 0.98 & 12.3 & 12.8  \\ 
J0217+0144 & OD 026&3FGL J0217.8+0143 & 1.715& \n & 2.36 & 1.64 & 0.62 & 13.1 & 13.5  \\ 
J0228+6721\tablenotemark{a} & 4C +67.05&\n & 0.523& 0.02 & 1.37 & 0.88 & 0.57 & 12.8 & 13.0  \\ 
J0231+1322 & 4C +13.14&\n & 2.059& 0.07 & 2.06 & 1.68 & 1.37 & 12.3 & 12.8  \\ 
J0237+2848 & 4C +28.07&3FGL J0237.9+2848 & 1.206& \n & 4.27 & 2.65 & 0.82 & 12.9 & 13.2  \\ 
J0238+1636 & AO 0235+164&3FGL J0238.6+1636 & 0.940& \n & 6.73 & 1.24 & 1.36 & 13.0 & 13.2  \\ 
J0241$-$0815 & NGC 1052&\n & 0.005037& 0.10 & 1.63 & 1.11 & 0.66 & \n & \n  \\ 
J0244+6228\tablenotemark{a} & TXS 0241+622&\n & 0.045& \n & 2.33 & 1.69 & 1.35 & 13.4 & 13.5  \\ 
J0303+4716\tablenotemark{a} & 4C +47.08&3FGL J0303.6+4716 & \n& 0.05 & 2.44 & 1.91 & 1.42 & 13.4 & \n  \\ 
J0319+4130 & 3C 84&3FGL J0319.8+4130 & 0.0176& 1.38 & 29.80 & 19.83 & 1.09 & 12.9 & 12.9  \\ 
J0336+3218 & NRAO 140&3FGL J0336.5+3210 & 1.259& \n & 3.12 & 2.51 & 1.12 & 13.0 & 13.4  \\ 
J0339$-$0146 & CTA 26&3FGL J0339.5$-$0146 & 0.852& \n & 3.85 & 2.53 & 1.12 & 12.7 & 13.0  \\ 
J0359+5057\tablenotemark{a} & NRAO 150&\tablenotemark{b} & 1.520& 0.13 & 16.05 & 11.73 & 1.22 & 12.4 & 12.8  \\ 
J0403+2600 & CTD 026&\n & 2.109& \n & 2.16 & 1.90 & 1.37 & 12.3 & 12.8  \\ 
J0405$-$1308 & PKS 0403$-$13&3FGL J0405.5$-$1307 & 0.571& 0.76 & 1.13 & 0.87 & 0.60 & 13.2 & 13.4  \\ 
J0418+3801\tablenotemark{a} & 3C 111&3FGL J0418.5+3813c & 0.0491& 0.90 & 6.16 & 3.13 & 0.89 & 13.3 & 13.3  \\ 
J0423$-$0120 & PKS 0420$-$01&3FGL J0423.2$-$0119 & 0.9161& \n & 8.65 & 5.32 & 0.63 & 12.8 & 13.1  \\ 
J0424+0036 & PKS 0422+00&3FGL J0424.7+0035 & 0.268& \n & 1.03 & 0.57 & 0.50 & 14.0 & 14.1  \\ 
J0433+0521 & 3C 120&\tablenotemark{c} & 0.033& 0.39 & 3.54 & 2.08 & 0.63 & 13.5 & 13.6  \\ 
J0442$-$0017 & NRAO 190&3FGL J0442.6$-$0017 & 0.845& \n & 2.30 & 1.46 & 1.36 & 13.0 & 13.3  \\ 
J0449+1121 & PKS 0446+11&3FGL J0449.0+1121 & 2.153& \n & 1.74 & 1.07 & 0.69 & 12.8 & 13.2  \\ 
J0453$-$2807 & OF $-$285&3FGL J0453.2$-$2808 & 2.559& \n & 1.72 & 1.55 & \n & 12.5 & 13.0  \\ 
J0457$-$2324 & PKS 0454$-$234&3FGL J0457.0$-$2324 & 1.003& \n & 2.39 & 1.83 & 1.01 & 12.7 & 13.0  \\ 
J0501$-$0159 & S3 0458$-$02&3FGL J0501.2$-$0157 & 2.286& 0.07 & 1.52 & 1.12 & 0.51 & 13.0 & 13.5  \\ 
J0530+1331 & PKS 0528+134&3FGL J0530.8+1330 & 2.070& \n & 3.54 & 1.83 & 0.33 & 12.8 & 13.3  \\ 
J0532+0732 & OG 050&3FGL J0532.7+0732 & 1.254& 0.06 & 1.65 & 1.35 & 0.94 & 12.6 & 12.9  \\ 
J0533+4822\tablenotemark{a} & TXS 0529+483&3FGL J0533.2+4822 & 1.160& 0.03 & 1.64 & 1.30 & 0.82 & 13.1 & 13.4  \\ 
J0541$-$0541 & PKS 0539$-$057&1FGL J0540.9$-$0547 & 0.838& 0.03 & 1.41 & 0.82 & 1.44 & 12.4 & 12.7  \\ 
J0555+3948\tablenotemark{a} & DA 193&\n & 2.363& 0.21 & 4.21 & 3.85 & 0.74 & \n & \n  \\ 
J0607$-$0834 & OC $-$010&3FGL J0608.0$-$0835 & 0.870& \n & 3.33 & 2.15 & 0.92 & 12.1 & 12.3  \\ 
J0609$-$1542 & PKS 0607$-$15&\n & 0.3226& \n & 5.99 & 3.64 & 0.52 & 12.2 & 12.3  \\ 
J0646+4451 & OH 471&\n & 3.396& \n & 3.81 & 3.35 & 0.86 & 11.8 & 12.5  \\ 
J0650$-$1637\tablenotemark{a} & PKS 0648$-$16&3FGL J0650.4$-$1636 & \n& 0.16 & 2.63 & 1.83 & 0.69 & \n & \n  \\ 
J0721+7120 & S5 0716+71&3FGL J0721.9+7120 & 0.127& 0.04 & 4.12 & 1.87 & 1.14 & 14.4 & 14.5  \\ 
J0725$-$0054\tablenotemark{a} & PKS 0723$-$008&3FGL J0725.8$-$0054 & 0.127& 0.05 & 4.55 & 4.12 & 2.19 & 13.4 & 13.5  \\ 
J0730$-$1141\tablenotemark{a} & PKS 0727$-$11&3FGL J0730.2$-$1141 & 1.591& \n & 9.54 & 6.86 & 1.34 & 12.6 & 13.0  \\ 
J0733+5022 & TXS 0730+504&3FGL J0733.8+5021 & 0.720& \n & 0.74 & 0.60 & 0.51 & 12.7 & 12.9  \\ 
J0738+1742 & OI 158&3FGL J0738.1+1741 & 0.450& 0.08 & 0.92 & 0.59 & 0.46 & 13.5 & 13.7  \\ 
J0739+0137 & OI 061&3FGL J0739.4+0137 & 0.1894& 0.03 & 2.41 & 1.21 & 0.75 & 13.2 & 13.3  \\ 
J0741+3112 & OI 363&\n & 0.631& 0.05 & 1.78 & 1.20 & 0.58 & \n & \n  \\ 
J0745+1011 & PKS B0742+103&\n & 2.624& 0.04 & 1.47 & 1.38 & 0.73 & \n & \n  \\ 
J0745$-$0044 & OI $-$072&\n & 0.996& \n & 1.55 & 1.37 & 0.80 & \n & \n  \\ 
J0748+2400 & S3 0745+24&3FGL J0748.3+2401 & 0.4092& \n & 1.63 & 1.17 & 1.05 & 12.8 & 13.0  \\ 
J0750+1231 & OI 280&3FGL J0750.6+1232 & 0.889& \n & 5.04 & 4.06 & 1.11 & 12.6 & 12.9  \\ 
J0757+0956 & PKS 0754+100&3FGL J0757.0+0956 & 0.266& \n & 2.01 & 1.46 & 0.69 & 13.4 & 13.5  \\ 
J0808+4950 & OJ 508&3FGL J0807.9+4946 & 1.436& \n & 1.03 & 0.47 & 0.43 & 12.1 & 12.5  \\ 
J0808$-$0751 & PKS 0805$-$07&3FGL J0808.2$-$0751 & 1.837& 0.03 & 2.53 & 1.52 & 1.53 & 12.8 & 13.2  \\ 
J0811+0146 & OJ 014&3FGL J0811.3+0146 & 1.148& \n & 1.36 & 0.97 & 0.86 & 13.0 & 13.3  \\ 
J0818+4222 & OJ 425&3FGL J0818.2+4223 & \n& \n & 2.12 & 1.55 & 1.17 & 13.1 & \n  \\ 
J0824+3916 & 4C +39.23&3FGL J0824.9+3916 & 1.216& 0.13 & 1.36 & 1.05 & 0.82 & 12.6 & 12.9  \\ 
J0825+0309 & PKS 0823+033&3FGL J0826.0+0307 & 0.505& \n & 2.14 & 0.76 & 1.02 & 13.0 & 13.2  \\ 
J0830+2410 & OJ 248&3FGL J0830.7+2408 & 0.942& \n & 2.08 & 1.31 & 0.94 & 12.6 & 12.8  \\ 
J0831+0429 & OJ 049&3FGL J0831.9+0430 & 0.174& \n & 1.20 & 0.66 & 0.70 & 13.6 & 13.6  \\ 
J0836$-$2016 & PKS 0834$-$20&3FGL J0836.5$-$2020 & 2.752& 0.14 & 2.18 & 2.07 & 0.63 & 11.9 & 12.4  \\ 
J0841+7053 & 4C +71.07&3FGL J0841.4+7053 & 2.218& 0.02 & 3.07 & 2.19 & 1.16 & 12.4 & 12.9  \\ 
J0840+1312 & 3C 207&3FGL J0840.8+1315 & 0.680& 0.13 & 1.79 & 1.27 & 0.93 & \n & \n  \\ 
J0850$-$1213 & PMN J0850$-$1213&3FGL J0850.2$-$1214 & 0.566& \n & 1.62 & 0.57 & 1.70 & 13.1 & 13.3  \\ 
J0854+2006 & OJ 287&3FGL J0854.8+2006 & 0.306& \n & 11.05 & 4.68 & 1.88 & 13.7 & 13.8  \\ 
J0902$-$1415 & PKS B0859$-$140&\n & 1.339& 0.16 & 1.28 & 0.83 & 0.65 & \n & \n  \\ 
J0909+0121 & 4C +01.24&3FGL J0909.1+0121 & 1.026& \n & 2.69 & 1.42 & 0.97 & 13.5 & 13.8  \\ 
J0920+4441 & S4 0917+44&3FGL J0920.9+4442 & 2.189& \n & 2.17 & 2.02 & 1.48 & 13.0 & 13.5  \\ 
J0921+6215 & OK 630&3FGL J0921.8+6215 & 1.453& 0.04 & 1.45 & 1.20 & 0.75 & 12.5 & 12.9  \\ 
J0927+3902 & 4C +39.25&\n & 0.695& \n & 12.33 & 10.39 & 0.89 & 12.4 & 12.6  \\ 
J0948+4039 & 4C +40.24&3FGL J0948.6+4041 & 1.249& 0.06 & 1.82 & 1.43 & 0.94 & 12.5 & 12.9  \\ 
J0958+6533 & S4 0954+65&3FGL J0958.6+6534 & 0.367& \n & 2.53 & 1.31 & 0.99 & 13.4 & 13.5  \\ 
J0958+4725 & OK 492&3FGL J0957.4+4728 & 1.882& \n & 1.20 & 0.75 & 0.63 & 12.2 & 12.6  \\ 
J1033+4116 & S4 1030+41&3FGL J1033.2+4116 & 1.117& \n & 2.71 & 1.39 & \n & 12.8 & 13.1  \\ 
J1037$-$2934 & PKS 1034$-$293&3FGL J1037.0$-$2934 & 0.312& 0.09 & 2.57 & 2.28 & 0.94 & 12.8 & 12.9  \\ 
J1038+0512 & PKS 1036+054&\n & 0.473& \n & 1.95 & 1.29 & 0.69 & 12.4 & 12.5  \\ 
J1041+0610 & 4C +06.41& \n & 1.265& \n & 1.53 & 1.34 & 0.83 & 12.7 & 13.0  \\ 
J1048$-$1909 & PKS 1045$-$18&\n & 0.595& \n & 2.09 & 1.21 & 1.20 & 12.7 & 12.9  \\ 
J1058+0133 & 4C +01.28&3FGL J1058.5+0133 & 0.888& 0.08 & 5.51 & 4.77 & 0.91 & 12.8 & 13.1  \\ 
J1127$-$1857 & PKS 1124$-$186&3FGL J1127.0$-$1857 & 1.048& 0.08 & 3.29 & 1.80 & 1.14 & 13.2 & 13.5  \\ 
J1130$-$1449 & PKS 1127$-$14&3FGL J1129.9$-$1446 & 1.184& 0.04 & 3.71 & 2.29 & 0.94 & 12.7 & 13.0  \\ 
J1130+3815 & B2 1128+38&3FGL J1131.4+3819 & 1.733& 0.03 & 1.52 & 1.23 & 0.98 & 12.3 & 12.7  \\ 
J1153+4931 & 4C +49.22&3FGL J1153.4+4932 & 0.3334& 0.04 & 1.93 & 1.78 & 1.76 & 13.1 & 13.2  \\ 
J1153+8058 & S5 1150+81&\n & 1.250& 0.06 & 1.11 & 0.99 & 0.62 & 12.4 & 12.8  \\ 
J1159+2914 & 4C +29.45&3FGL J1159.5+2914 & 0.725& 0.10 & 3.60 & 2.26 & 0.64 & 12.9 & 13.2  \\ 
J1215$-$1731 & PKS 1213$-$17&\n & \n& 0.05 & 1.95 & 1.70 & 0.75 & \n & \n  \\ 
J1222+0413 & 4C +04.42&3FGL J1222.4+0414 & 0.966& \n & 1.42 & 1.00 & 1.21 & 12.8 & 13.1  \\ 
J1224+2122 & 4C +21.35&3FGL J1224.9+2122 & 0.434& 0.16 & 2.56 & 1.99 & 1.05 & 13.1 & 13.3  \\ 
J1229+0203 & 3C 273&3FGL J1229.1+0202 & 0.1583& 5.07 & 27.60 & 22.97 & 0.63 & 13.9 & 14.0  \\ 
J1230+1223 & M87&3FGL J1230.9+1224 & 0.00436& 24.07 & 4.62 & 2.13 & 1.02 & \n & \n  \\ 
J1256$-$0547 & 3C 279&3FGL J1256.1$-$0547 & 0.536& \n & 33.71 & 15.90 & 1.05 & 12.9 & 13.1  \\ 
J1310+3220 & OP 313&3FGL J1310.6+3222 & 0.997& \n & 3.06 & 2.70 & 0.77 & 13.0 & 13.3  \\ 
J1327+2210 & B2 1324+22&3FGL J1326.8+2211 & 1.403& \n & 1.46 & 1.03 & 0.64 & 12.5 & 12.9  \\ 
J1337$-$1257 & PKS 1335$-$127&3FGL J1337.6$-$1257 & 0.539& 0.09 & 6.80 & 4.17 & 0.58 & 12.5 & 12.7  \\ 
J1408$-$0752 & PKS B1406$-$076&3FGL J1408.8$-$0751 & 1.494& \n & 1.10 & 0.95 & 0.69 & 12.7 & 13.1  \\ 
J1415+1320 & PKS B1413+135&3FGL J1416.0+1325 & 0.247& \n & 1.23 & 0.67 & 0.51 & 12.8 & 12.9  \\ 
J1419+3821 & B3 1417+385&3FGL J1419.8+3819 & 1.831& \n & 1.21 & 0.62 & 0.73 & 12.4 & 12.9  \\ 
J1436+6336 & VIPS 0792&\n & 2.066& 0.07 & 1.54 & 1.43 & 1.09 & 12.6 & 13.1  \\ 
J1459+7140 & 3C 309.1&\n & 0.904& 0.80 & 0.80 & 0.61 & 0.41 & \n & \n  \\ 
J1504+1029 & OR 103&3FGL J1504.4+1029 & 1.8385& \n & 3.33 & 1.51 & 1.22 & 12.7 & 13.2  \\ 
J1507$-$1652 & PKS 1504$-$167&\n & 0.876& 0.04 & 1.04 & 0.86 & 0.45 & 12.4 & 12.7  \\ 
J1512$-$0905 & PKS 1510$-$08&3FGL J1512.8$-$0906 & 0.360& 0.04 & 6.68 & 2.50 & 1.34 & 13.5 & 13.7  \\ 
J1516+1932 & PKS 1514+197&3FGL J1516.9+1926 & 1.070& \n & 1.75 & 1.02 & 1.71 & 12.9 & 13.2  \\ 
J1517$-$2422 & AP Librae&3FGL J1517.6$-$2422 & 0.049& \n & 2.58 & 2.12 & 0.75 & 14.2 & 14.2  \\ 
J1522$-$2730 & PKS 1519$-$273&3FGL J1522.6$-$2730 & 1.297& \n & 0.99 & 0.87 & 0.51 & 12.8 & 13.2  \\ 
J1540+1447 & 4C +14.60&3FGL J1540.8+1449 & 0.606& 0.12 & 1.18 & 0.92 & 0.83 & 13.1 & 13.3  \\ 
J1549+0237 & PKS 1546+027&3FGL J1549.4+0237 & 0.414& \n & 2.60 & 1.79 & 0.58 & 12.9 & 13.1  \\ 
J1550+0527 & 4C +05.64&3FGL J1550.5+0526 & 1.417& 0.07 & 3.04 & 2.74 & 0.93 & 12.8 & 13.2  \\ 
J1608+1029 & 4C +10.45&3FGL J1608.6+1029 & 1.232& \n & 1.45 & 0.88 & 0.63 & 12.9 & 13.2  \\ 
J1613+3412 & DA 406&3FGL J1613.8+3410 & 1.40& \n & 3.76 & 2.57 & 0.66 & 12.3 & 12.6  \\ 
J1625$-$2527 & PKS 1622$-$253&3FGL J1625.7$-$2527 & 0.786& 0.19 & 2.91 & 2.00 & 0.67 & 12.6 & 12.9  \\ 
J1626$-$2951 & PKS 1622$-$29&3FGL J1626.0$-$2951 & 0.815& \n & 2.38 & 1.46 & 0.61 & 12.6 & 12.9  \\ 
J1635+3808 & 4C +38.41&3FGL J1635.2+3809 & 1.813& \n & 4.15 & 3.37 & 0.93 & 12.5 & 12.9  \\ 
J1638+5720 & OS 562&3FGL J1637.9+5719 & 0.751& 0.04 & 2.33 & 1.46 & 0.89 & \n & \n  \\ 
J1640+3946 & NRAO 512&3FGL J1640.6+3945 & 1.666& \n & 1.85 & 1.09 & 1.08 & 12.1 & 12.5  \\ 
J1642+3948 & 3C 345&3FGL J1642.9+3950 & 0.593& 0.30 & 9.15 & 7.36 & 0.74 & 13.0 & 13.2  \\ 
J1642+6856 & 4C +69.21&\n & 0.751& \n & 4.73 & 2.46 & 2.17 & 12.4 & 12.7  \\ 
J1658+0741 & PKS 1655+077&\n & 0.621& 0.04 & 2.10 & 1.88 & 0.94 & 12.4 & 12.6  \\ 
J1727+4530 & S4 1726+45&3FGL J1727.1+4531 & 0.717& \n & 2.49 & 1.27 & 1.14 & 12.8 & 13.1  \\ 
J1733$-$1304 & NRAO 530&3FGL J1733.0$-$1305 & 0.902& 0.46 & 5.26 & 4.26 & 0.35 & 13.0 & 13.3  \\ 
J1740+5211 & 4C +51.37&3FGL J1740.3+5211 & 1.379& \n & 2.00 & 1.27 & 0.75 & 13.1 & 13.5  \\ 
J1743$-$0350 & PKS 1741$-$03&3FGL J1744.3$-$0353 & 1.054& \n & 5.66 & 3.72 & 0.72 & 12.3 & 12.7  \\ 
J1751+0939 & 4C +09.57&3FGL J1751.5+0939 & 0.322& \n & 7.37 & 4.38 & 0.84 & 13.0 & 13.1  \\ 
J1753+2848 & B2 1751+28&\n & 1.118& \n & 1.88 & 1.07 & 0.94 & \n & \n  \\ 
J1800+3848 & B3 1758+388B&\n & 2.092& \n & 1.29 & 1.06 & 0.72 & 12.1 & 12.6  \\ 
J1801+4404 & S4 1800+44&3FGL J1801.5+4403 & 0.663& 0.05 & 1.46 & 0.96 & 0.84 & 12.6 & 12.8  \\ 
J1800+7828 & S5 1803+784&3FGL J1800.5+7827 & 0.6797& 0.02 & 3.40 & 2.65 & 1.04 & 13.4 & 13.7  \\ 
J1806+6949 & 3C 371&3FGL J1806.7+6949 & 0.051& 0.23 & 1.64 & 1.34 & 0.95 & 14.2 & 14.2  \\ 
J1824+5651 & 4C +56.27&3FGL J1824.2+5649 & 0.664& 0.09 & 1.67 & 1.50 & 0.64 & 13.2 & 13.4  \\ 
J1829+4844 & 3C 380&3FGL J1829.6+4844 & 0.692& 1.12 & 2.56 & 2.10 & 0.96 & 13.0 & 13.2  \\ 
J1842+6809 & S4 1842+68&3FGL J1842.8+6810 & 0.472& 0.02 & 1.64 & 0.60 & 0.92 & 12.3 & 12.4  \\ 
J1849+6705 & S4 1849+67&3FGL J1849.2+6705 & 0.657& \n & 3.13 & 2.44 & 1.03 & 13.2 & 13.4  \\ 
J1911$-$2006 & PKS B1908$-$201&3FGL J1911.2$-$2006 & 1.119& \n & 3.66 & 1.65 & 1.29 & 12.9 & 13.3  \\ 
J1923$-$2104 & OV $-$235&3FGL J1923.5$-$2104 & 0.874& \n & 2.07 & 1.75 & 1.09 & \n & \n  \\ 
J1924$-$2914 & PKS B1921$-$293&3FGL J1924.8$-$2914 & 0.3526& 1.39 & 20.99 & 13.40 & 0.86 & 12.5 & 12.6  \\ 
J1925+2106\tablenotemark{a} & PKS B1923+210&\n & \n& 0.05 & 2.33 & 1.82 & 0.70 & 12.2 & \n  \\ 
J1927+7358 & 4C +73.18&\n & 0.302& 0.03 & 4.59 & 4.05 & 1.02 & 13.2 & 13.3  \\ 
J1939$-$1525 & PKS 1936$-$15&\n & 1.657& 0.03 & 0.88 & 0.69 & 0.50 & 12.0 & 12.4  \\ 
J1959+4044\tablenotemark{a} & Cygnus A&\n & 0.0561& 66.10 & 29.01 & 1.11 & 1.01 & \n & \n  \\ 
J2000$-$1748 & PKS 1958$-$179&3FGL J2001.0$-$1750 & 0.652& \n & 3.29 & 1.26 & 1.16 & 12.6 & 12.8  \\ 
J2007+4029\tablenotemark{a} & TXS 2005+403&\n & 1.736& \n & 3.88 & 3.49 & 1.25 & 12.2 & 12.6  \\ 
J2005+7752 & S5 2007+77&3FGL J2005.2+7752 & 0.342& 0.08 & 1.55 & 0.80 & 0.83 & 13.4 & 13.6  \\ 
J2011$-$1546 & PKS 2008$-$159&\n & 1.180& \n & 2.48 & 2.15 & 1.01 & 12.7 & 13.0  \\ 
J2015+3710\tablenotemark{a} & TXS 2013+370&3FGL J2015.6+3709 & 0.859& 0.16 & 4.42 & 2.95 & 0.99 & \n & \n  \\ 
J2023+3153\tablenotemark{a} & 4C +31.56&3FGL J2023.2+3154 & 0.356& 0.14 & 1.27 & 1.19 & 0.59 & \n & \n  \\ 
J2022+6136 & OW 637&\n & 0.227& \n & 2.41 & 2.32 & 0.75 & \n & \n  \\ 
J2025+3343\tablenotemark{a} & B2 2023+33&3FGL J2025.2+3340 & 0.219& \n & 6.42 & 3.71 & 1.95 & \n & \n  \\ 
J2031+1219 & PKS 2029+121&3FGL J2031.8+1223 & 1.213& \n & 2.10 & 1.51 & 1.53 & 12.6 & 13.0  \\ 
J2038+5119\tablenotemark{a} & 3C 418&3FGL J2038.8+5113 & 1.686& 0.30 & 3.43 & 3.12 & 1.24 & 12.5 & 12.9  \\ 
J2123+0535 & OX 036&3FGL J2123.6+0533 & 1.941& \n & 1.99 & 1.64 & 0.53 & 12.5 & 13.0  \\ 
J2131$-$1207 & PKS 2128$-$12&\n & 0.501& 0.12 & 2.22 & 1.81 & 0.67 & \n & \n  \\ 
J2134$-$0153 & 4C $-$02.81&3FGL J2134.1$-$0152 & 1.284& \n & 2.72 & 2.25 & 1.02 & 13.0 & 13.3  \\ 
J2136+0041 & PKS 2134+004&\n & 1.932& \n & 7.32 & 6.86 & 0.99 & \n & \n  \\ 
J2139+1423 & OX 161&\n & 2.427& \n & 2.95 & 2.63 & 1.03 & 12.2 & 12.7  \\ 
J2148+0657 & 4C +06.69&1FGL J2148.5+0654 & 0.999& 0.04 & 5.90 & 5.39 & 0.52 & 12.6 & 12.9  \\ 
J2158$-$1501 & PKS 2155$-$152&3FGL J2158.0$-$1501 & 0.672& \n & 3.99 & 1.88 & 1.52 & 12.8 & 13.1  \\ 
J2202+4216 & BL Lac&3FGL J2202.7+4217 & 0.0686& 0.02 & 7.83 & 4.40 & 1.24 & 13.7 & 13.7  \\ 
J2203+1725 & PKS 2201+171&3FGL J2203.4+1725 & 1.076& \n & 1.66 & 1.15 & 0.83 & 13.2 & 13.5  \\ 
J2203+3145 & 4C +31.63&3FGL J2203.7+3143 & 0.2947& \n & 3.54 & 2.90 & 0.94 & 13.9 & 14.0  \\ 
J2212+2355 & PKS 2209+236&3FGL J2212.0+2355 & 1.125& \n & 1.28 & 0.91 & 0.67 & 12.1 & 12.4  \\ 
J2218$-$0335 & PKS 2216$-$03&\n & 0.901& 0.11 & 2.04 & 1.51 & 0.60 & 12.3 & 12.6  \\ 
J2225+2118 & DA 580&\n & 1.959& 0.09 & 1.91 & 1.37 & 1.54 & 12.5 & 12.9  \\ 
J2225$-$0457 & 3C 446&3FGL J2225.8$-$0454 & 1.404& 0.88 & 8.05 & 5.86 & 1.01 & 12.9 & 13.3  \\ 
J2229$-$0832 & PHL 5225&3FGL J2229.7$-$0833 & 1.5595& \n & 3.80 & 2.39 & 1.76 & 12.8 & 13.2  \\ 
J2232+1143 & CTA 102&3FGL J2232.5+1143 & 1.037& 0.14 & 5.91 & 3.45 & 0.90 & 12.4 & 12.7  \\ 
J2236+2828 & CTD 135&3FGL J2236.3+2829 & 0.790& \n & 1.53 & 1.19 & 1.04 & 12.9 & 13.2  \\ 
J2246$-$1206 & PKS 2243$-$123&\n & 0.632& \n & 2.75 & 2.33 & 0.79 & 12.3 & 12.6  \\ 
J2253+1608 & 3C 454.3&3FGL J2254.0+1608 & 0.859& \n & 28.12 & 13.10 & 1.50 & 13.4 & 13.7  \\ 
J2258$-$2758 & PKS 2255$-$282&3FGL J2258.0$-$2759 & 0.927& \n & 4.72 & 3.13 & 0.51 & 12.6 & 12.9  \\ 
J2327+0940 & OZ 042&3FGL J2327.7+0941 & 1.841& \n & 2.82 & 1.37 & 1.25 & 12.2 & 12.7  \\ 
J2334+0736 & TXS 2331+073&3FGL J2334.1+0732 & 0.401& \n & 1.54 & 1.07 & 1.01 & 12.7 & 12.9  \\ 
J2348$-$1631 & PKS 2345$-$16&3FGL J2348.0$-$1630 & 0.576& \n & 2.48 & 1.96 & 0.85 & 12.9 & 13.1  \\ 
J2354+4553 & 4C +45.51&3FGL J2354.1+4605 & 1.986& 0.15 & 1.01 & 0.81 & 0.44 & 12.2 & 12.7  \\ 
\enddata

\tablecomments{Columns are as follows: 
(1) J2000 name, 
(2) other name,
(3) Fermi catalog name,
(4) redshift,
(5) arcsecond-scale flux density at 15 GHz in Jy,
(6) maximum 15 GHz VLBA flux density in Jy during 3FGL time window,
(7) median 15 GHz VLBA flux density in Jy during 3FGL time window,
(8) radio activity index, 
(9) log of observed synchrotron peak frequency in Hz (observer frame),
(10) log of observed synchrotron peak frequency in Hz (rest frame).}

\tablenotetext{a}{ {Low} galactic latitude ($|b| < 10^\circ$)}
\tablenotetext{b}{{\it Fermi} LAT detection reported by \cite{2014ATel.5838....1C}}
\tablenotetext{c}{{\it Fermi} LAT detection reported by \cite{2010ApJ...720..912A}}
\end{deluxetable*} 

\vskip 1truein
\begin{deluxetable*}{lcc} 
\tablecolumns{3} 
\tabletypesize{\scriptsize} 
\tablewidth{0pt}  
\tablecaption{\label{stattests} Statistical Tests on LAT and Non-LAT
Sub-samples}  
\tablehead{\colhead{Property} & 
\colhead{Kolmogorov-Smirnov} & \colhead{Wilcoxon Rank Sum} 
 }      

\startdata 
Synchrotron peak location (observed)            & $1\times10^{-4}$&$3 \times 10^{-5}$ \\
Synchrotron peak location (rest frame)          & $1\times10^{-4}$&$1 \times 10^{-5}$ \\
Maximum observed jet speed                      & $2\times10^{-6}$&$3 \times 10^{-5}$ \\
15 GHz modulation index                         & $5\times10^{-4}$&$9 \times 10^{-5}$ \\
Redshift                                        & 0.50 & 0.45 \\
Radio activity index\tablenotemark{a}           & 0.10 & 0.14 \\
Median VLBA radio flux density\tablenotemark{a} & 0.86 & 0.50 \\
Maximum VLBA radio flux density\tablenotemark{a}& 0.04 & 0.05 \\
Median VLBA radio luminosity\tablenotemark{a}   & 0.86 & 0.80 \\
Maximum VLBA radio luminosity\tablenotemark{a}  & 0.76 & 0.73 \\
\enddata

\tablecomments{The tabulated values indicate the probability of
  obtaining the observed distributions in the LAT and non-LAT detected
  sub-samples, under the null hypothesis that they came
  from the same parent distribution.}

\tablenotetext{a}{During the 3FGL time window.}
\end{deluxetable*}

\end{document}